\documentclass[10pt,conference]{IEEEtran} 
\usepackage[utf8]{inputenc}
\usepackage{cite}
\usepackage{graphicx}
\usepackage{hyperref}
\usepackage{multirow}
\usepackage{tabulary}
\usepackage{xcolor}

\newcommand{\keyword}[1]{\textit{#1}}
\newcommand{\code}[1]{\textsl{{#1}}}

\title{An Alternative to Cells for Selective Execution\\of Data Science Pipelines}

\author{\IEEEauthorblockN{Lars Reimann}
\IEEEauthorblockA{Institute of Computer Science III\\
University of Bonn, Germany\\
Email: reimann@cs.uni-bonn.de}
\and
\IEEEauthorblockN{Günter Kniesel-Wünsche}
\IEEEauthorblockA{Institute of Computer Science III\\
University of Bonn, Germany\\
Email: gk@cs.uni-bonn.de}
}

\begin{document}

\IEEEoverridecommandlockouts
\IEEEpubid{
  \begin{minipage}{\textwidth}
  \noindent\rule[0.5ex]{0.5\textwidth}{0.75pt}\\
  Accepted at the 45th Int. Conf on Software Engineering, NIER Track\\
  ICSE-NIER’23, May 14–20, 2023, Melbourne, Australia\\
  \copyright2023 IEEE 
  \end{minipage}
}

\maketitle

\IEEEpubidadjcol

\begin{abstract}
Data Scientists often use notebooks to develop Data Science (DS) pipelines, particularly since they allow to selectively execute parts of the pipeline. However, notebooks for DS have many well-known flaws. We focus on the following ones in this paper: (1) Notebooks can become littered with code cells that are not part of the main DS pipeline but exist solely to make decisions (e.g. listing the columns of a tabular dataset). (2) While users are allowed to execute cells in any order, not every ordering is correct, because a cell can depend on declarations from other cells. (3) After making changes to a cell, this cell and all cells that depend on changed declarations must be rerun. (4) Changes to external values necessitate partial re-execution of the notebook. (5) Since cells are the smallest unit of execution, code that is unaffected by changes, can inadvertently be re-executed.

% \discussionbefore{Im nächsten Absatz würde ich nicht mehr nur von Zellen sprechen, sondern erst mal betonen, dass wir nicht nur Zellen, sondern Notebooks an und für sich ersetzen wollen, und zwar durch eine herkömmliche IDE, die aber Strukturierungsmechanismen unterstützt die Zellen entsprechen und die Verknüpfung  von Text und ausführbarem Code, die der Wesentliche Grund für den Erfolg von Notebooks für Unterrichtszwecke ist.}
%
To solve these issues, we propose to replace cells as the basis for the selective execution of DS pipelines. Instead, we suggest populating a context-menu for variables with actions fitting their type (like listing columns if the variable is a tabular dataset). These actions are executed based on a data-flow analysis to ensure dependencies between variables are respected and results are updated properly after changes. Our solution separates pipeline code from decision making code and automates dependency management, thus reducing clutter and the risk of making errors. 

%
% \discussionafter{Vorschlag: We show that we can avoid all these issues by using traditional IDEs extended with the features that made notebooks so popular. In our approach, developers only need to write the real DS pipeline code while the IDE supports decision making via context menu items that allow inspection of data values. Execution of the entire pipeline, or until a certain statement of the pipeline, is available as a context menu item, too. What needs to be executed is automatically determined by static analysis, which is possible because the pipeline code is not littered with additional value inspection code. Leveraging our clean separation of concerns between DS pipeline code and value inspection, we support structuring parts of the pipeline into functions as an equivalent of notebook cells and provide the ability to add arbitrary text to each function as an equivalent of the instructional text that made notebooks so popular. Our approach also makes versioning of pipeline code easy, because only pipeline code is edited in the first place. So stored versions are not polluted with unnecessary code.}

\end{abstract}


\section{Introduction}
\label{sec:introduction}
 Notebooks allow users to 
%%% - split their code into cells, which  
%%% +
write code in cells that
can be executed independently, in any order.
Execution results, such as visualizations, are commonly shown close to the code cells that produced them. Code cells can be interspersed with text cells, to offer explanations or document decisions, which follows the paradigm of literate programming \cite{knuthLiterateProgramming1984}. 
%Overall, this makes notebooks well suited to (1) develop interactive tutorials and (2) rapidly explore different approaches to solve a given problem. Because of (2),
Overall, this makes notebooks well suited for the explorative development process of Data Science (DS) pipelines \cite{biswasArtPracticeData2022}. Here, developers often write small pieces of code, run them, and use the results to change existing code or decide what to implement next.

Various flavors of notebooks exist, with Jupyter Notebook \cite{kluyverJupyterNotebooksPublishing} being most popular according to the 2021 Kaggle Survey on DS \cite{kaggle2021KaggleMachine}. Jupyter Notebook requires language-specific \keyword{kernels} to execute code and to power IDE features 
like auto-completion, but is otherwise language-agnostic. Since Jupyter Notebook is a web application, it can run locally or be hosted as a service, like Google Colaboratory \cite{googleGoogleColaboratory}, Kaggle \cite{kaggleNotebooksDocumentation}, or Amazon SageMaker \cite{amazonUseAmazonSageMaker}. Some integrated development environments (IDEs) like PyCharm \cite{jetbrainsJupyterNotebookSupport} or Visual Studio Code \cite{kabirNotebooksVisualStudio} incorporate Jupyter Notebook. JupyterLab \cite{grangerJupyterLabBuildingBlocks} is eventually meant to replace the default Jupyter Notebook GUI. The issues we discuss in this paper are independent from a specific notebook variant, however, since they stem from the core concept of notebooks: Cells.

\begin{figure}[ht]
    \centering
    \includegraphics[width=.40\textwidth]{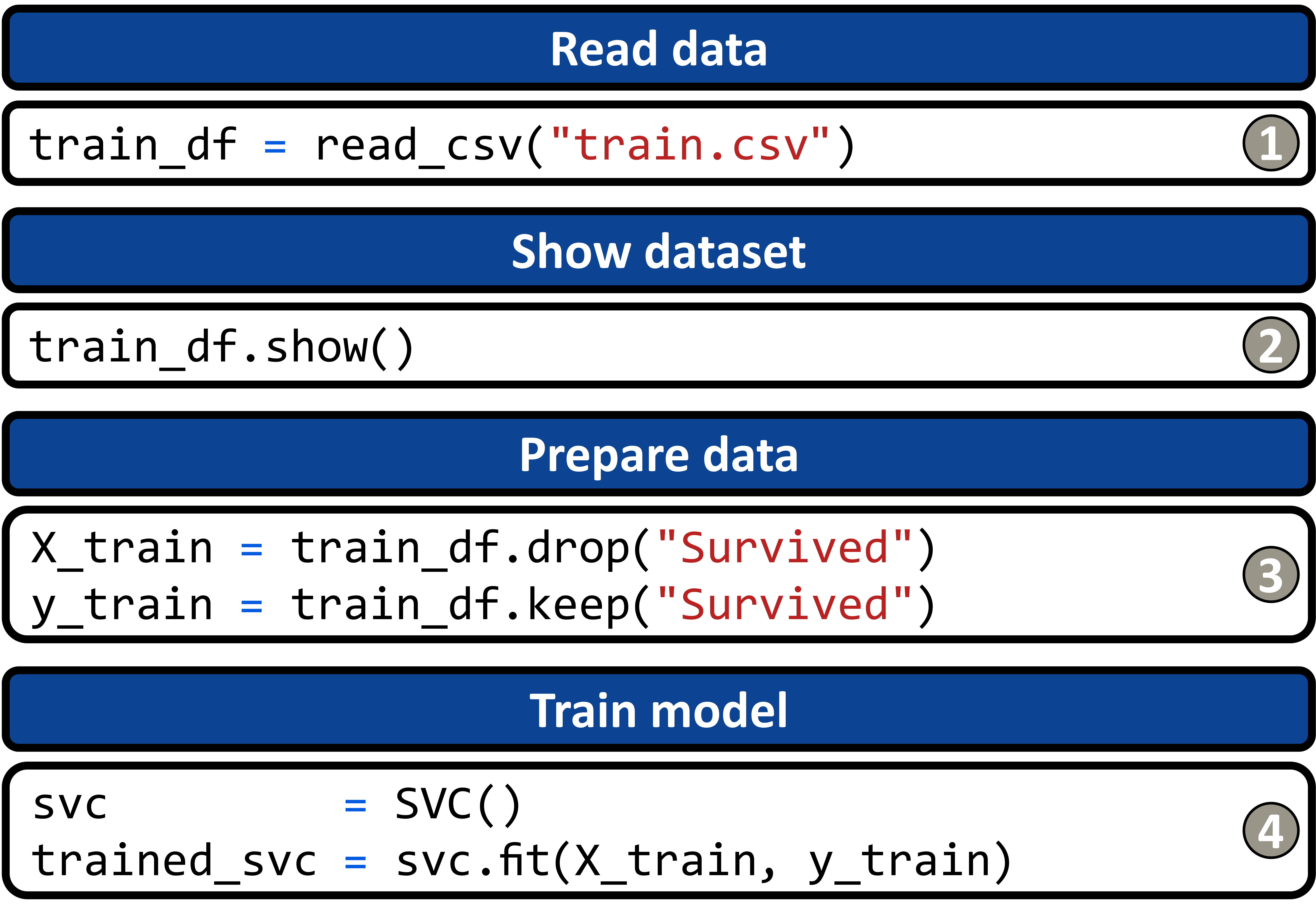}
    \caption{Example for a typical cell structure of a DS notebook with text cells (blue) and code cells (white). 
    %%% + 
    Result cells are omitted. 
    }
    \label{fig:example_notebook}
\end{figure}

Fig. \ref{fig:example_notebook} shows an example of the typical cell structure of a notebook for DS\footnote{Based on a popular notebook from Kaggle (\url{https://www.kaggle.com/code/startupsci/titanic-data-science-solutions}).} using Python as the programming language. Text cells (blue) and code cells (white) alternate. The code cells (1) read a CSV file containing the training data, (2) view the dataset to gain a general understanding and to know which attributes it has, (3) separate feature vector and target, (4) configure a model (support vector machine for classification), and train it\footnote{The \code{fit} method in this example is supposed to return a new trained model rather than mutate the untrained model stored in \code{svc}.}. For the sake of brevity, we omit result cells.

Based on Fig. \ref{fig:example_notebook} we can illustrate the problems we want to discuss in this paper:

\begin{enumerate}
    \item Actual pipeline code (Cells 1, 3, 4) is mixed with value inspection code (Cell 2). The value inspection code exists solely to write more pipeline code (here Cell 3, which needs the name of the target column). Keeping value inspection cells in the notebook afterwards increases clutter \cite{keryStoryNotebookExploratory2018, ruleExplorationExplanationComputational2018} and negatively impacts performance, since they might get re-executed unnecessarily.
    %\discussionafter{Wir könnten hinzufügen: This clutter also reduces the benefit of approaches to automatic versioning of DS pipelines \cite{***}, since versions will contain code that will later inhibit understanding the differences between versions.}
    \item Users can execute cells in any order. This manipulates the internal state of the notebook (e.g. contained variables and their assigned values), which is maintained throughout the entire session. However, not all orderings are 
    %%% - legal 
    %%% +
    correct
    \cite{grusDonNotebooks, keryStoryNotebookExploratory2018, headManagingMessesComputational2019, pimentelLargeScaleStudyQuality2019, chattopadhyayWhatWrongComputational2020, mackeFinegrainedLineageSafer2021, pimentelUnderstandingImprovingQuality2021}: For example, executing Cell 3 first on a blank state is erroneous, since this cell depends on \code{train\_df}, which is computed in Cell 1.
    \item 
    %- The notebook state becomes stale, after code changes. This affects the internal state as well as any results cells. 
    %+
    Code changes partially invalidate the internal notebook state and results cells \cite{grusDonNotebooks, keryStoryNotebookExploratory2018, headManagingMessesComputational2019, pimentelLargeScaleStudyQuality2019, chattopadhyayWhatWrongComputational2020, mackeFinegrainedLineageSafer2021, pimentelUnderstandingImprovingQuality2021}.
    Say, after executing the entire notebook from Fig. \ref{fig:example_notebook}, we add more data preparation steps to Cell 3, 
    before assigning \code{X\_train} and \code{y\_train}. 
    Now, the values of \code{X\_train} and \code{y\_train} in the internal notebook state are outdated, so Cell 3 must be rerun. However, \code{trained\_svc}, which depends on these values, is also outdated, so Cell 4 must be rerun, too. In large notebooks, keeping track of all cells that must be rerun gets complicated, leading to developers frequently rerunning the entire notebook to be safe \cite{chattopadhyayWhatWrongComputational2020},
    %%% +
    wasting computing and development time.
    \item Even without code changes, notebook state can become stale: For example, the dataset we read from disk in Cell 1 might change. Like in Problem 3, the notebook needs to be partially re-executed to account for such an event.
    \item Since cells are the smallest unit of execution, even code that is unaffected by changes is re-executed. In the example from Problem 3, we rerun Cell 4 completely, although the value of \code{svc} %in the internal state of the notebook 
    is still valid. Calling the \code{SVC} constructor is fast, but for the same reason long-running data preparation or model training operations might be rerun unnecessarily, drastically slowing the feedback loop. This leads to an ``expand then reduce'' pattern \cite{keryStoryNotebookExploratory2018}, where developers first write small code cells, which can be executed independently, to iterate quickly and later combine them into bigger cells.
    %%% ? Dazu noch ein Satz, warum das keine ausreichende Lösung ist? ---> Man kann nicht beliebig fein granular werden, denn sonst geht die Strukturierung verloren und man muss sehr viele kleine Zellen selbst überblicken und nachträglich neu ausführen.
    This defeats the purpose of cells as a means to group logically related code together.
\end{enumerate}

\section{State of the Art}
\label{sec:state-of-the-art}

A natural solution to ensure code gets executed in the correct order (Problem 2) and gets re-executed after changes (Problem 3) is to derive dependencies between cells using data-flow analysis and run a cell only after all cells it depends on:
%
%The idea to use data-flow analysis to ensure that code gets executed in the correct order (RQ 2) and gets re-executed after changes (RQ 3) is not new. 
%
Dataflow notebooks \cite{koopDataflowNotebooksEncoding} assign a unique identifier to cells, which stays the same even as the code in the cell changes. These IDs are used to describe the dependencies between cells. When a cell gets executed, the system ensures that all upstream dependencies are available already. 
% It also offers capabilities to update downstream cells.
Nodebook \cite{NodebookStitchFix} keeps track of inputs and outputs of cells. Inputs are determined by parsing the code in the cell, while outputs are discovered by comparing the internal state of the notebook after the cell was run to the state before. \cite{headManagingMessesComputational2019} describes how data-flow analysis can be used to create a polished version of a notebook that only contains the code needed to produce the results that the user selected. NBSAFETY \cite{mackeFinegrainedLineageSafer2021} uses data-flow analysis to detect unsafe interactions with the cells in a notebook and offer resolution advisories. To achieve this, NBSAFETY uses a mix of dynamic and static analysis. ReSplit \cite{titovReSplitImprovingStructure2022} analyzes definition-usage chains between cells and within the same cell and then suggests an alternative mapping of code to cells to ensure that tightly coupled code resides in the same cell. 

%%% +
However, none of the existing approaches solve Problems 1, 4, and 5: They do not handle changes to values outside the notebook (Problem 4), use complete cells as the smallest unit of execution thus failing to avoid re-execution of parts unaffected by changes (Problem 5), and do not even try to address the tangling of pipeline and value inspection code (Problem 1).   

%%% - Alles folgende ersetzen durch vorangegengene Neuformulierung:
%
%However, none of the existing approaches take changes to values outside the notebook into account (Problem 4). Moreover, they still use cells as the basis for execution.
%rather than replacing them entirely for this purpose and keeping them only as a structuring element.
%This fails to address Problem 5 since 
%%%% - without careful management of cells, 
%code might get rerun unnecessarily. For Problem 1 we did not find an existing solution. 

% \paragraph*{Version management} Besides mixing value inspection and pipeline code, keeping different versions of pipeline code in the same notebook is another source of clutter \cite{keryStoryNotebookExploratory2018}. Variolite \cite{keryVarioliteSupportingExploratory2017} is a tool for the versioning of small pieces of code that is specifically designed for exploratory tasks like the implementation of DS pipelines. \cite{headManagingMessesComputational2019} suggests to create a version browser for results and the code that created them. The experiment tracking of MLflow \cite{chenDevelopmentsMLflowSystem2020} provides a means to implement such a version browser. \cite{weinmanForkItSupporting2021} proposes to show alternatives directly in the layout of notebooks as different parallel paths.

\section{Research Questions}
\label{sec:research-questions}

This brings us to the following research questions:

\begin{itemize}
    \item \textbf{RQ 1}: How can we separate pipeline code and value inspection code?
    \item \textbf{RQ 2}: How can we ensure changes to external values trigger re-execution of code that depends on them?
    \item \textbf{RQ 3}: How can we avoid unnecessarily executing code?
\end{itemize}

RQ 1 aims to reduce clutter, RQ 2 is about correctness and RQ 3 addresses performance.

\section{Approach Overview}
\label{sec:approach-overview}

To solve these issues, we propose to keep cells only as a means to connect code to related results and instructional or documenting text, but abandon them as the basis for the selective execution of DS pipelines. Instead, we suggest to

\begin{enumerate}
    \item 
    %%% + 
    address RQ 1 by introducing different roles for code cells (Sec. \ref{sec:tags-for-code-cells}), or by statically typing variables and letting the development environment dynamically populate a context-menu for variables with actions that can be run on a variable of the respective type (Sec. \ref{sec:introspection}),
    \item address RQ 2 and RQ 3 by using static code analysis 
    % Folgendes später ergänzen:
    % data-flow analysis, information about the purity of functions, and information about cached values 
    to derive a correct and minimal \emph{execution plan} for a selected action (Sec. \ref{sec:data-flow-analysis}). 
    %%% ? Das bezieht sich auf die actions und klingt damit so, als ob man es nicht bräuchte, wenn man kein actions hat sondern einfach nur Zellen taggt. Wie können wir das klarer formulieren?
\end{enumerate}

%\begin{enumerate}
%    \item  introduce different roles for code cells (Sec. \ref{sec:tags-for-code-cells}),
%    \item use static typing to decide which actions can be run on a variable and allow users to select one of those actions in a context-menu (Sec. \ref{sec:introspection}),
%    \item use static code analysis 
    % Folgendes später ergänzen:
    % data-flow analysis, information about the purity of functions, and information about cached values 
%    to derive a correct and minimal execution plan for a selected action (Sec. \ref{sec:data-flow-analysis}). 
    %%% ? Das bezieht sich auf die actions und klingt damit so, als ob man es nicht bräuchte, wenn man kein actions hat sondern einfach nur Zellen taggt. Wie können wir das klarer formulieren?
%\end{enumerate}

An execution plan is \keyword{correct}, if it contains all operations that must be executed and guarantees that each operation is executed only when all its input values are up-to-date, including external values (RQ 2). It is \keyword{minimal} (RQ 3), if it contains no operations that would only affect already up-to-date values or would compute values that are not accessed by other operations in the graph.
% \discussionafter{Korrektheit wird teils durch den Graph und teils durch die datenflussbasierte Ausführungsstrategie garantiert. Wie und an welcher Stelle kann man das am besten ausdrücken?}. 

\section{Tags for Code Cells}
\label{sec:tags-for-code-cells}

For teaching materials \cite{deprattiJupyterNotebooksTextbook2020} or documentation, keeping value inspection cells in the notebook can be helpful, to outline each step of the development process of a DS pipeline, including decision-making \cite{keryStoryNotebookExploratory2018}. However, value inspection code is then scattered across the entire notebook, creating clutter, and re-executed each time the notebook is run, wasting time.

\begin{figure}[ht]
    \centering
    \includegraphics[width=.38\textwidth]{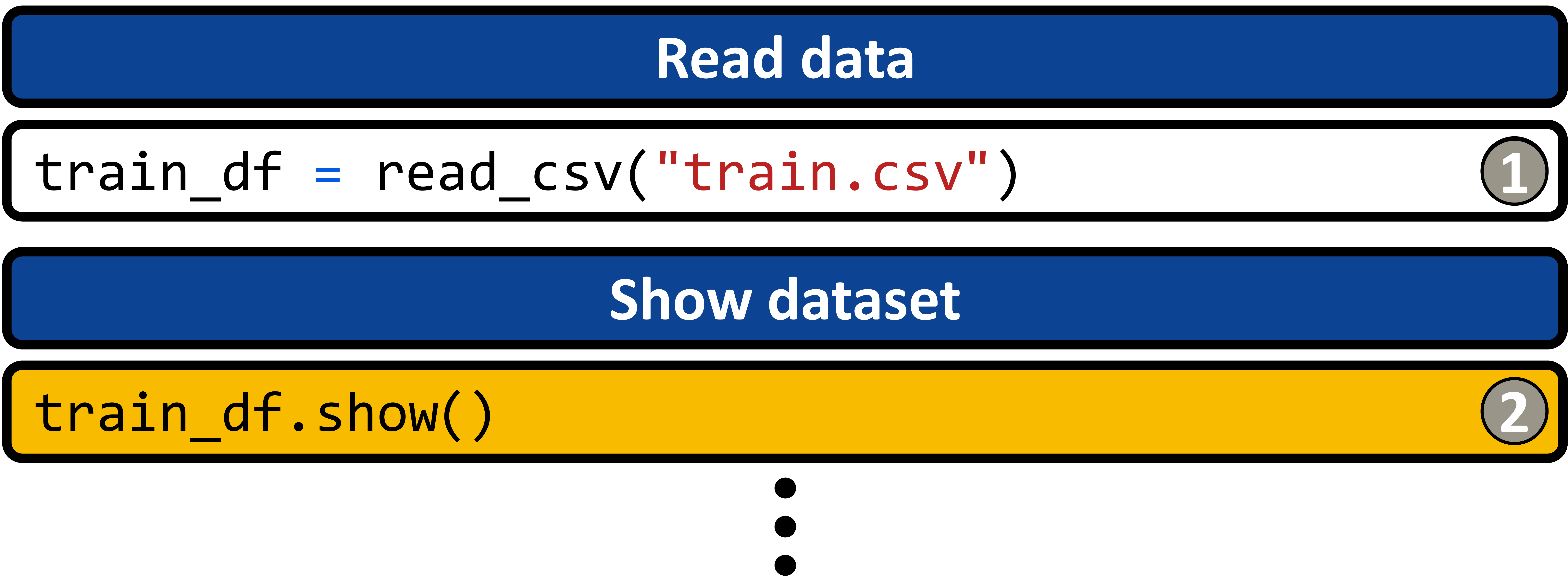}
    \caption{Distinguishing pipeline code cells (white), and value inspection code cells (yellow) at the start of the notebook from Fig. \ref{fig:example_notebook}.}

%  \caption{Start of the notebook from Fig. \ref{fig:example_notebook} with text cells (blue), pipeline code cells (white), and value inspection code cells (yellow).}
\label{fig:notebook_with_roles}
\end{figure}

To 
%%% - still 
separate pipeline code and value inspection code (RQ 1) in this case, we suggest to 
%%% - introduce \emph{tags} for code cells, which describe their purpose.
%%% +
tag code cells by their purpose, as ``pipeline'' or ``inspection''.
Fig. \ref{fig:notebook_with_roles} shows this for the start of the notebook from Fig. \ref{fig:example_notebook} with pipeline cells (white) and 
%%% - value 
inspection cells (yellow). With a filter, a user can then narrow down the cells and show, say, just the pipeline cells. Only cells that are currently shown 
%%% - should get 
are 
executed. 
%%% - Text cells associated with code cells can similarly be tagged and filtered.
%%% +
Text cells can be tagged and filtered in the same way.
% ODER
% Each text cell is tagged and filtered like its associated code cell.

% Text cells that correspond to hidden code cells should also be hidden. For this, we can either allow them to be tagged as well, use a heuristic and assume text cells correspond to the code cell below, or introduce a more rigid structure using ``supercells'' that bundle a text cell, a code cell, and a result cell. Then the entire supercell can be tagged.

\section{Context-Sensitive Inspection of Values}
\label{sec:introspection}

Outside of educational 
%%% - DS 
notebooks, value inspections cells are often executed only once,
%%% - to write correct code afterwards. 
%%% +
to decide what to do next.
For example, after running Cell 2 in Fig. \ref{fig:example_notebook} we know that the target attribute is called ``Survived'' and can use this knowledge to write Cell 3. Leaving Cell 2 in the notebook has little 
%%% - effect on documentation 
%%% + 
documentation value, 
since we already manifested the extracted information in Cell 3.

\begin{figure}[t]
    \centering
    \includegraphics[width=.38\textwidth]{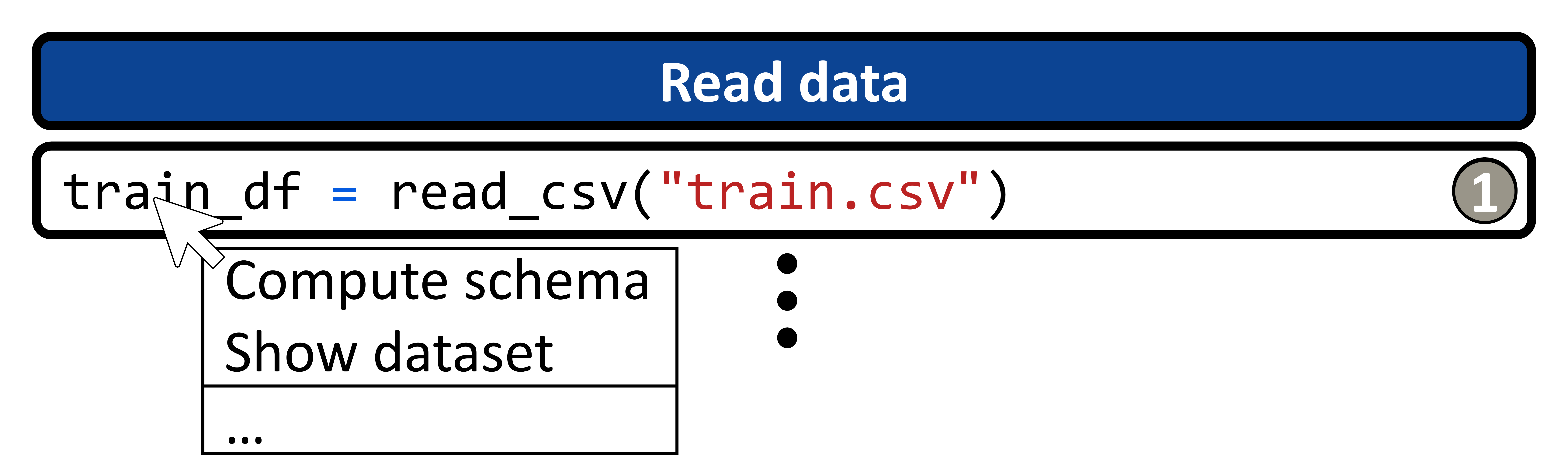}
    \caption{
    %%% - Hiding value inspection actions in a context-menu.
    %%% +
    Replacing value inspection cells by type-specific inspection actions in the context-menu of variables. 
    }
    \label{fig:notebook_with_context_menu}
\end{figure}

%Outside of educational DS notebooks, value inspection cells get executed once and can be removed afterwards.
%%% + 
In contexts where inspection cells have no documentation effect,
we suggest to avoid writing 
%%% - these value inspection cells
them
in the first place, by additionally offering 
%%% value 
inspection actions in a context-menu for variables (Fig. \ref{fig:notebook_with_context_menu}). This way, notebooks can only contain pipeline code if desired, 
%%% + 
completely eliminating the tangling with
%%% - value 
inspection code (RQ 1).

To know which actions can be triggered on a variable, we need to know its type. Ideally, the type should be known statically, so we do not need to run code to determine which actions can be triggered on a variable. Moreover, the type should be inferred, so users can concentrate on writing their pipeline code as they are used to.
%%% ? Hier müsste noch etwas dazu gesagt werden, wie wir uns vorstellen, dass man statische Typen inferieren kann - und auf unser anderes Paper verweisen. Sonst sind das aussagen im luftleeren Raum, man weiss nicht was wir damit meinen. 

% \discussionafter{Das argument kann ich leider nicht nachvollziehen. Siehe folgenden Absatz:}
% \neu{In notebooks for not statically typed languages, such as Python, determining the type of a variable would only be possible after running the code that computes its value. This would would be no problem, since the value can only be inspected anyway after it is computed.}

\begin{figure}[t]
    \centering
    \includegraphics[width=.38\textwidth]{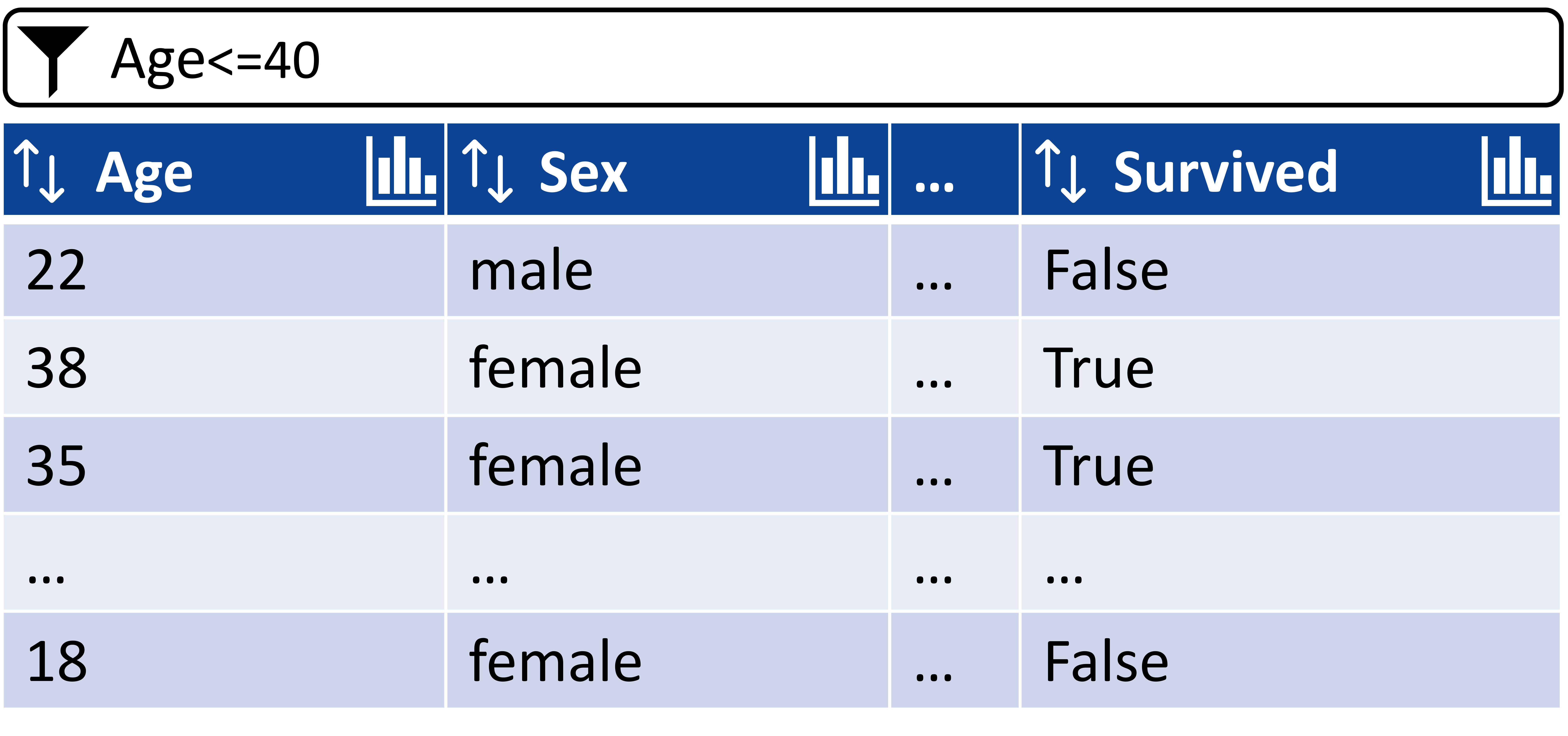}
    \caption{Mockup for the interactive output of the ``Show dataset'' action. Users can filter the table (funnel icon), sort it by a column (arrow icons), generate histograms (chart icons), or trigger other 
    %%% +
    inspection and analysis actions 
    from this view without having to 
    %%% - go back to the context menu or 
    %%% +
    write any code.}
    \label{fig:table_mockup}
\end{figure}

When a user selects an action, code 
%%% + 
that implements the action
%%% - is generated and <-- kann doch auch schon vorher statisch implementiert sein
is executed in the background (see Sec. \ref{sec:data-flow-analysis} for details). Results of value inspection actions can be 
%%% - 
closed after inspection or
kept outside the main flow of the notebook entirely
%%% - as well, for example 
in separate tabs, windows, or Sticky Cells \cite{wangStickyLandBreakingLinear2022}
%%%- , which 
that
float on top of the notebook and maintain their position even when the notebook is scrolled. Fig. \ref{fig:table_mockup} shows a mockup of the potential output for the ``Show dataset'' action
%%% +
from Fig. \ref{fig:notebook_with_context_menu}. 
The data is displayed in an interactive table that the user can filter (funnel icon) or sort (arrow icons). Additional actions can be triggered directly from this view, without the need to go back to the context menu, e.g. for generating a histogram of a column (chart icons).

% To ensure that all values that the generated code depends on are computed already, we can build \alt{an execution} \neu{a dependency} graph using data-flow analysis (Sec. \ref{sec:data-flow-analysis}). Since the execution results are used to inform design choices in other parts of the notebook, it makes sense to show them in Sticky Cells \cite{wangStickyLandBreakingLinear2022} \discussionafter{Klingt im Prinzip schön. Aber eine ganze Tabelle in einer Sticky Cell??? WIe gehen wir mit großen Ergebnissen um?}. These float on top of the notebook and maintain their position even when the notebook is scrolled. The data-flow analysis can also be used to automatically update results when code is changed. \discussionafter{Dieser Satz nimmt den nächsten Abschnitt vorweg. Wenn er dir an dieser Stelle wichtig ist, einen Querverweis auf nächsten Abschnitt einfügen. } Once the results are no longer needed\discussionafter{When would that be?}, the Sticky Cells can simply be closed.

%\section{Data-Flow Analysis for Execution Plans}
%\label{sec:data-flow-analysis}

\section{Minimal Execution}
\label{sec:data-flow-analysis}

\begin{figure}[t]
    \centering
    \includegraphics[width=.38\textwidth]{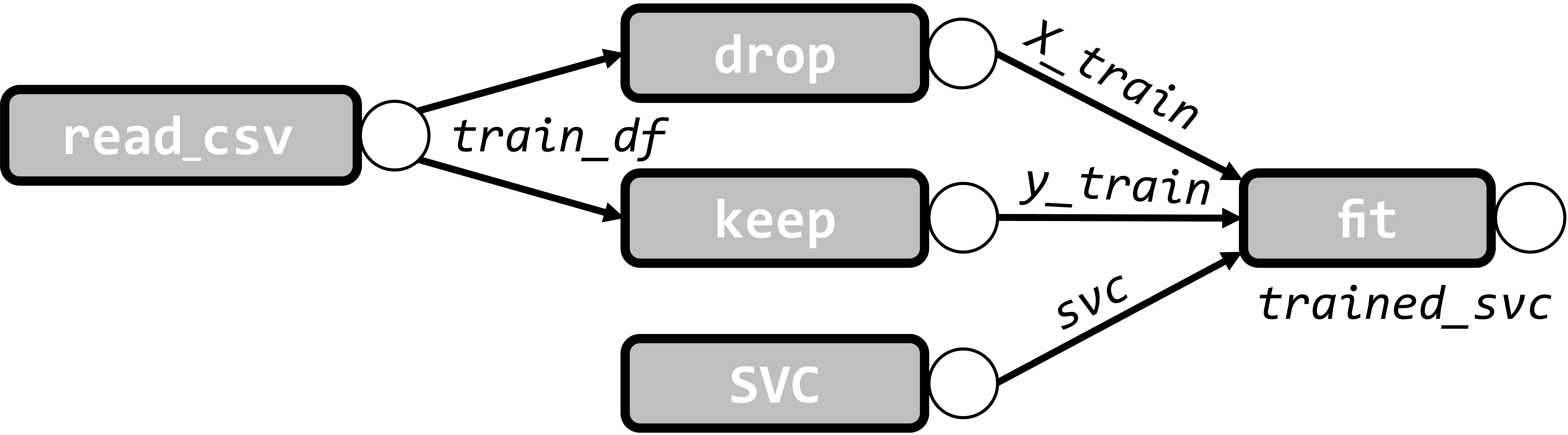}
    %{img/data_flow_graph.pdf}
    \caption{Data-flow graph for pipeline code from Fig. \ref{fig:example_notebook}. Rounded boxes represent operations and edges represent data-flow. Each edge is labelled by the name of the variable through which the respective data flows.}
    \label{fig:data_flow_graph}
\end{figure}

\begin{figure}[t]
    \centering
    \includegraphics[width=.253\textwidth]{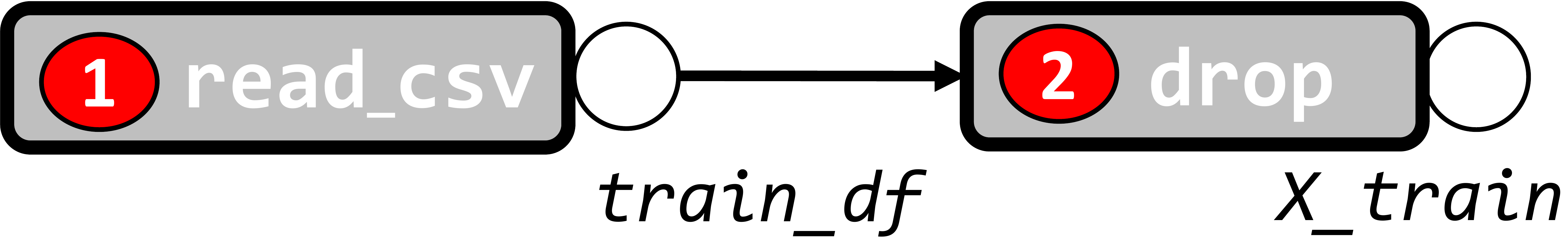}
    \caption{Execution plan to view the \code{X\_train} dataset. The numbers in the boxes indicate the implicit order of the operations.}
    \label{fig:execution_graph_selective}
\end{figure}

\paragraph*{Data-flow} As described in Sec. \ref{sec:introspection}, we want to % be able to 
trigger context-sensitive actions on \emph{variables} without executing unnecessary code (RQ 3). For example, if the user selects the ``Show dataset'' action on the variable \code{X\_train} from Fig. \ref{fig:example_notebook}, it is a waste of time to execute the entire Cell 3, since the value of \code{y\_train} is never used. This requires a fine-grained data-flow graph that focuses on individual operations rather than cells. Fig. \ref{fig:data_flow_graph} shows the data-flow graph for the example notebook from Fig. \ref{fig:example_notebook}. From this graph we can derive that we only need to evaluate the calls of \code{read\_csv} and \code{drop} to compute \code{X\_train}, leading to the simple execution plan in Fig. \ref{fig:execution_graph_selective}

\begin{figure}[ht]
    \centering
    \includegraphics[width=.38\textwidth]{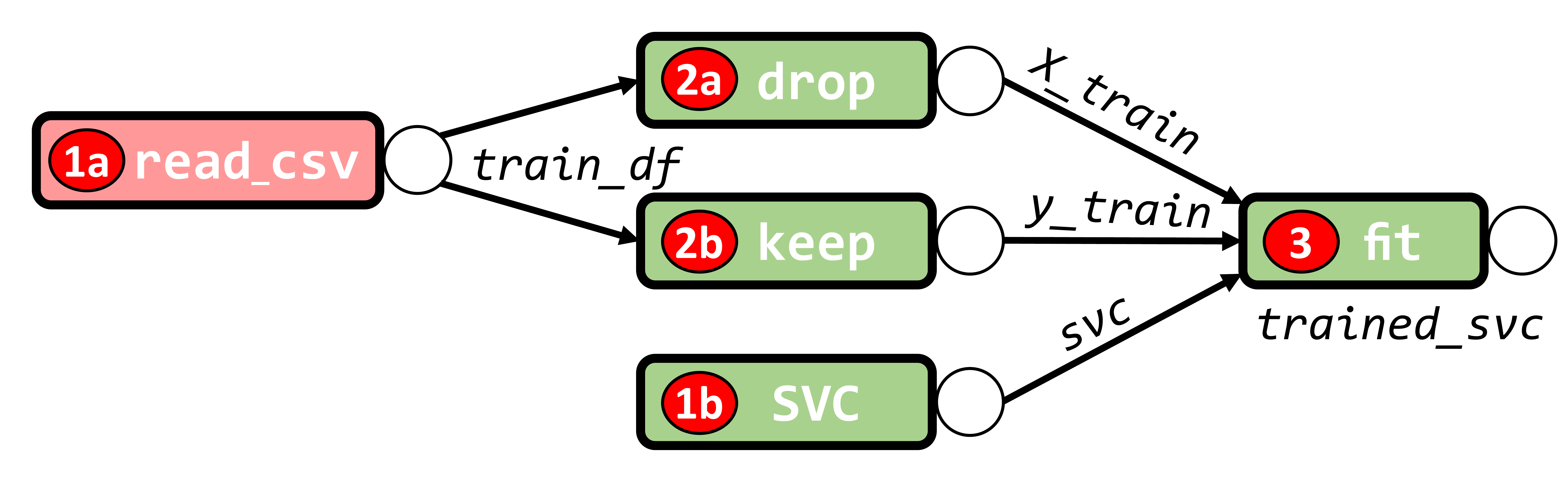}
    \caption{Extended data-flow graph for initial run of the notebook. Numbers indicate sequential execution, letters indicate potential parallelism.}
    \label{fig:execution_graph_initial}
\end{figure}

\paragraph*{Purity} We can derive an execution plan for 
the entire notebook up to the \code{fit} call, based on 
the data-flow graph from Fig. \ref{fig:data_flow_graph}.
The graph tells us that we need to run the entire notebook, since \code{fit} depends on all other operations. However, we can further optimize, if we know which operations are \keyword{pure}. A pure operations has no side effects and its outputs only depend on its parameters captured in the data-flow graph. Impure operations may read or modify external state (files), or global state (global variables or object attributes). 
% Two operations that do not depend on each other can be executed \emph{in parallel} if at least one is pure. 
Any pure operation can be executed \emph{in parallel} to any other \emph{independent} operation (that is, an operation on a different path of the dataflow graph). In contrast, independent impure operations must be executed according to their textual order. For this, additional edges that reflect the textual order are added between independent impure operations, yielding the complete execution plan. 
In our example, all operations are pure, except \code{read\_csv}, which reads a file from disk and is impure because the file contents might change although the value of the read operation's \code{path} parameter stays the same. Adding purity information (green background) and impurity information (red background) leads to the extended data-flow graph shown in Fig. \ref{fig:execution_graph_initial}. From it we can deduce that the calls to \code{read\_csv} and \code{SVC} can be run in parallel, as well as the calls to \code{drop} and \code{keep}. 
If the calls to \code{drop} and \code{keep} where impure and the call to \code{drop} occurred textually before the one to \code{keep}, an additional edge would point in Fig. \ref{fig:execution_graph_initial} from \code{drop} to \code{keep}.  

\paragraph*{Execution} 
The extended dataflow graph serves as an execution plan for computing the value of some variable \code{x}, when the user triggers an action (Sec. \ref{sec:introspection}) on  \code{x}:

% When the user triggers an action (Sec. \ref{sec:introspection}) on some variable \code{x}, the value of \code{x} is computed as follows:
\begin{enumerate}
    \item Build the dataflow graph.
    \item Extend it by purity information for operations.
    \item Extend it by textual order edges between independent impure operations.
    \item 
    % Build the backward slice \cite{slicing} of variable \code{x}, that is, 
    Eliminate operations that have no path to \code{x}. 
    \item
    Start executing (in any order or in parallel) nodes that have no incoming edges. After execution, delete the respective node and its outgoing edges.
    \item Repeat from Step 5, until the graph is empty, at which point the value of \code{x} is available. 
%\end{itemize}
\end{enumerate}
The first two steps ensure correctness. The first ensures that data dependencies are respected. The second preserves the order of impure operations. The third step removes all operations that have no effect on the value we want to compute, a first contribution towards minimality (RQ 3). 

% \begin{figure}[h]
%     \centering
%     \includegraphics[width=.40\textwidth]{img/data_flow_graph_with_purity.pdf}
%     \caption{Extended version of Fig. \ref{fig:data_flow_graph} with pure functions marked in green}
%     \label{fig:data_flow_graph_with_purity}
% \end{figure}

\paragraph*{Re-Execution} Let us now assume the initial execution plan from Fig. \ref{fig:execution_graph_initial} has been run, so the internal notebook state contains the current values of all variables. Let us further assume that the user has subsequently edited the call of the \code{drop} operation, to remove additional attributes from the feature vector after inspecting \code{X\_train}. Now, if the user requests the system to update the state of the notebook, we want to ensure that all code that is affected by the change gets rerun, without re-executing code unnecessarily (RQ 3). As with any rerun, we also need to take into account potential changes to external state, which can affect impure operations (RQ 2).

\begin{figure}[ht]
    \centering
    \includegraphics[width=.38\textwidth]{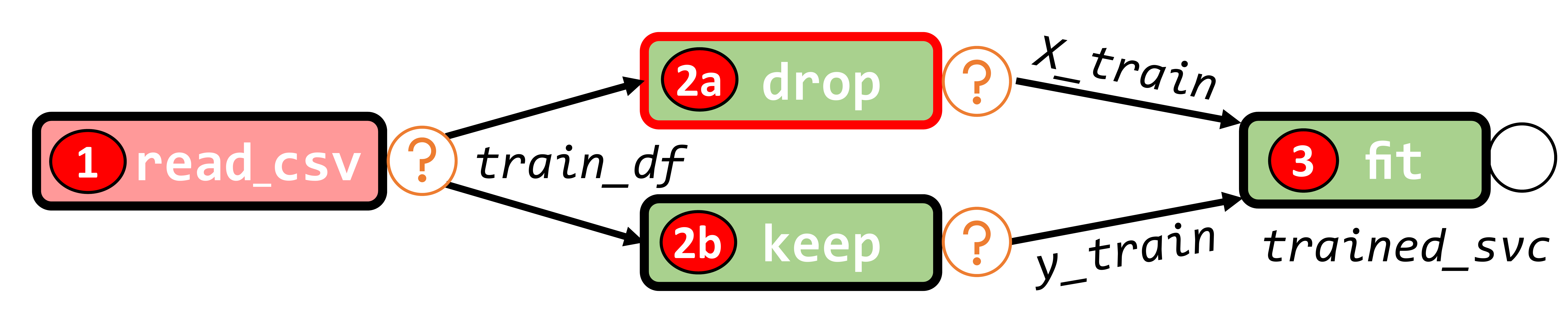}
    %{img/data_flow_graph_with_changes.pdf}
    \caption{%Extended version of Fig. \ref{fig:execution_graph_initial} 
    Re-execution plan with question marks indicating potentially stale values and red numbered labels indicating re-execution order of operations.}
    \label{fig:data_flow_graph_with_changes}
\end{figure}

Fig. \ref{fig:data_flow_graph_with_changes} illustrates the affected part of the notebook, after editing the \code{drop} call. The changed operation is marked by a red border. \keyword{Potentially stale} variables are indicated by yellow question marks:
\begin{itemize}
    \item \code{X\_train}, the output of the edited \code{drop} call,
    \item \code{train\_df}, the output of the impure operation \code{read\_csv} -- because the data-flow graph does not show the hidden dependencies of impure operations, we must always assume that something might have changed,
    \item
    \code{y\_train}, the output of the call to \code{keep} that depends on the potentially stale value \code{train\_df}.
\end{itemize}
All potentially stale values are indicated by a  question mark in Fig. \ref{fig:data_flow_graph_with_changes}. 
The call to SVC is unaffected by the change and its output is still  \keyword{up-to-date}. Therefore they are not included in the re-execution plan. 
%in Fig. \ref{fig:data_flow_graph_with_changes}.
% Other outputs are still \keyword{up-to-date}, shown by a green check mark in Fig. \ref{fig:data_flow_graph_with_changes}.

%\begin{figure}[ht]
%    \centering
%    \includegraphics[width=.38\textwidth]{img/data_flow_graph_with_changes.pdf}
%    \caption{Extended version of Fig. \ref{fig:data_flow_graph} with changed operations marked in red, pure operations marked in green, impure operations marked in grey, and information about staleness of cached values for variables (cross $\rightarrow$ known stale, question mark $\rightarrow$ potentially stale, check mark $\rightarrow$ up-to-date).}
%    \label{fig:data_flow_graph_with_changes}
%\end{figure}

The red labels in Fig. \ref{fig:data_flow_graph_with_changes} reflect the assumption that we must always rerun edited operations, impure operations, and pure operations with at least one potentially stale argument. Potentially stale arguments must be recomputed before they are used, which leads to the shown ordering.

\begin{figure}[ht]
    \centering
    \includegraphics[width=.38\textwidth]{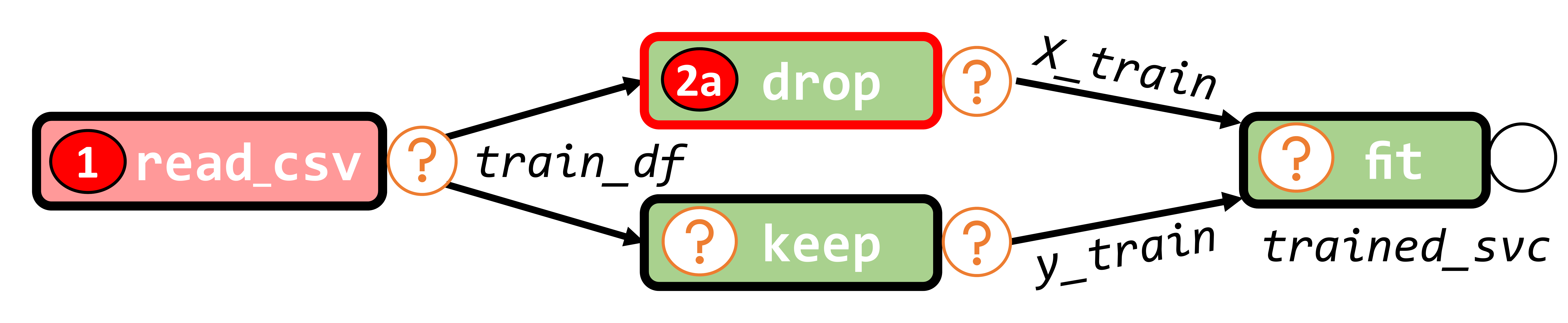}
    \caption{Re-execution plan with dynamic non-staleness checks. The question marks on an operation indicate that it is executed only if the dynamic non-staleness check fails for one of its potentially stale input values.}
    \label{fig:execution_graph_changed}
\end{figure}

A possibly more efficient approach is illustrated in Fig. \ref{fig:execution_graph_changed}. The yellow question marks on the calls to \code{keep} and \code{fit} indicate that these operations do not necessarily need re-execution, if recomputing their \emph{potentially} stale arguments produces the same values as before. For this \keyword{non-staleness check}, we can store the previous value and compare it to the new one. If it didn't change, we mark the argument as up-to-date. If all inputs of a pure operation are up-to-date, we do not need to re-run it and can mark its output as up-to-date.  
%Purity information can still be used to derive potential for parallelism. 
In the plan from Fig. \ref{fig:execution_graph_changed} the operations \code{drop} and \code{keep} can be run in parallel after \code{read\_csv}. If the dataset returned by \code{read\_csv} is unchanged, we can skip the call to \code{keep} and can mark \code{y\_train} as up-to-date.

This approach can eliminate expensive re-executions at the cost of remembering previous values and comparing them to new ones. For large datasets or models, such comparisons can be costly, too. Thus, the efficiency of the approach critically depends on the trade-off between re-execution time and comparison time. Even expensive comparisons might pay off, because they can eliminate re-execution of many downstream pure operations, including the followup checking of each of their results.

In this regard, an impure operation like \code{read\_csv} is a particularly worthwhile target for optimization: (1) It usually occurs right at the start of a pipeline, so many other operations depend on it and must be repeated if \code{read\_csv} yields new results. (2) The returned data is large, so an equality check is extremely expensive. 

We can fortunately avoid executing file operations subject to a simple check and the storing of minimal extra information: The time of the last modification of each accessed file. If we call \code{read\_csv} with the same path argument and the respective file has still the same last modified time, the operation returns the same result. We can, therefore, treat the last modified time internally as an extra, hidden argument, and the modified \code{read\_csv} operation as pure. This principle can be generalized to other cases, e.g. a random number generator that has a fixed seed is no source of impurity anymore. This way, some often occurring cases of impurity can be eliminated, accounting for noticeable re-execution speedups.  
It  allows us to treat all operations shown in Fig. \ref{fig:example_notebook} as pure.

%\neu{The structural equality checks for referentially distinct dataframes can then include a shortcut to return \code{True} if their names and timestamps are equal.
% }

%\footnote{We can also give advanced users the option to trade guaranteed correctness for performance by treating impure operations like pure operations and potentially stale values like up-to-date values.}.

\section{Technical Challenges and Solutions}
\label{sec:challengesAndSolutions}

The concept of notebooks builds on simple language-specific kernels. Our proposal requires additional kernel features, namely static typing for the context menu (Sec. \ref{sec:introspection}), data-flow analysis to derive a correct but minimal execution plan (Sec. \ref{sec:data-flow-analysis}), and purity inference as a basis for exploiting parallelism. %and memoization. 
Implementing our proposal for arbitrary languages, would require significant additional work from kernel developers. For some languages, it would even be impossible to implement our extensions fully. Python, for instance, allows dynamically importing modules 
%with the function 
via \code{importlib.import\_module(name)} \cite{pythonImports}. 
%An adversary user could load the module name from an encrypted file, obfuscating the module that gets loaded and preventing its static analysis. 
Reading the module name from an encrypted file would prevent static analysis of the module. 

Instead of building on an existing language, we developed Safe-DS \cite{safe-ds}, a domain specific language for implementing DS pipelines. Safe-DS can integrate existing Python libraries via a simple stub language in a statically type-safe way. This lets the compiler infer the type information that we need to provide context-specific actions on variables. The design of Safe-DS as a language for creating pipelines rather than implementing the algorithms used in the pipeline makes data-flow analysis easy, due to the lack of loops, recursion, and conditionals, which guarantees an acyclic data-flow graph. Safe-DS additionally offers annotations to mark operations as pure, empowering the compiler to prune the execution plan even in cases where purity cannot be inferred. Lastly, Safe-DS has a standard library that prefers pure operations.
%, which can be memoized.

% We overcome such limitations by separating the language issues from the IDE issues addressed in this paper. The purity-based memoization and the data-flow based execution semantics are implemented as extensions of Safe-DS\cite{safe-ds}, %\footnote{\url{https://github.com/Safe-DS/DSL}}
% a domain specific language (DSL) for developing DS pipelines.  
% Safe-DS is a good basis because it is able to integrate existing DS libraries via a simple stub language in a statically type-safe way, including libraries written in highly dynamic languages, like Python. Its static type system includes schemata of tabular datasets and effects of data operations on schemata. This lets the compiler infer the type information that we need to provide context-specific actions on variables. The design of Safe-DS as a language for creating pipelines rather than implementing the algorithms used in the pipeline makes data-flow analysis easy, due to the lack of loops, recursion, and conditionals, which guarantees an acyclic data-flow graph. Safe-DS additionally offers annotations to mark functions as pure, empowering the compiler to prune the execution plan even in cases where purity cannot be inferred. Last but not least, Safe-DS has a standard library that prefers pure functions, which can be memoized. 
%, and (4) has a standard library that prefers pure functions, which can be memoized. 
%for DS that is able to integrate the rich ecosystem of existing DS libraries. 

\section{Limitations}
\label{sec:limitations}

Generalizing our approach to notebooks written in arbitrary languages can by tricky. It depends on the quality of static analysis for types, data-flow and purity available for the respective language. 
%Clearly, not every analysis is feasible. However, what is feasible is often sufficient for practical purposes. 
%Development of better analysis tools is always an option. 
If at least data-flow analysis works, one can alternatively annotate libraries with types and purity information manually. But for non-trivial libraries, this may quickly exceed available time. However, one can use a combined approach that performs a static analysis and then lets developers review and complement its results by annotating library elements 
% Hier die EIschränkung auf Pirity und Types
with non-inferrable type or purity information \cite{reimannImprovingLearnabilityMachine2022}. Such combined tools limit the amount of manual annotation, thus making it possible to exploit the full potential of our approach in spite of the limits of static analysis.

\section{Future Plans}
\label{sec:future-plans}

\paragraph*{Prototype} We currently extend Safe-DS\footnote{\url{https://github.com/Safe-DS/DSL}} by the discussed computation of execution graphs based on data-flow analysis and purity information. An experimental IDE for Safe-DS is already available as an extension for Visual Studio Code\footnote{\url{https://marketplace.visualstudio.com/items?itemName=safe-ds.safe-ds}}. After evolving Safe-DS, we will extend its IDE by the proposed, type-based context menu for inspection of variables, including presentation of tabular datasets (Fig. \ref{fig:table_mockup}), to eliminate inspection code from the pipeline. 
%Our GitHub repository
%\footnote{\url{https://github.com/Safe-DS/DSL}} also includes the current Safe-DS language version and an initial fraction of an API that integrates the main functionality of widely-used DS libraries, such as pandas, scikit-learn, matplotlib, seaborn, etc. Using the API-Editor tool \cite{reimannImprovingLearnabilityMachine2022}, we wrap the implementation of these libraries into a small, consistent, redundancy-free API that should eventually evolve into and easy to start with entry point API for beginners in applied data science. Taken together, the improvements at API, language and IDE level are intended to reduce the costs of DS development, from initial API learning to rapid and error-free every-day use. 

%We continue to iteratively extend the API, language and IDE to the point where we feel it is able to express a sufficiently large set of application scenarios from Kaggle. 

\paragraph*{Usability study} 
Once the prototype is complete, it  will be compared quantitatively and qualitatively to Jupyter Notebook. We will 
%focus on the potential impact on code quality 
measure effects on code comprehension 
(by separating value inspection and pipeline code), correctness (by enforcing execution of code in the right order and re-execution of changed code), and performance (by avoiding re-execution of unchanged code).

\paragraph*{Purity Inference for Python}
%Moreover, we plan to add purity inference to the API-Editor \cite{reimannImprovingLearnabilityMachine2022}, a tool to automatically create wrappers that implement a changed API around a Python library. It also creates Safe-DS stubs that reflect the inferred information, which lets us create Safe-DS purity annotations if we can prove that a function is pure.

% 
Moreover, we plan to add purity inference (pure, impure, unknown) to the API-Editor \cite{reimannImprovingLearnabilityMachine2022}, use it to provide the unknown information, and let purity information be reflected in the Safe-DS stubs that it creates. We will also use its ability to automatically create wrappers that implement a uniform DS API on the basis of existing Python libraries. 

\section{Conclusions}
\label{sec:conclusions}

In this paper, we presented an alternative to cells as a basis to selectively execute parts of a DS pipeline. We use static code analysis to derive a fine-grained data-flow graph that focuses on individual operations rather than cells (Sec. \ref{sec:data-flow-analysis}). From this, we derive an execution plan that is \keyword{correct} (contains all operations and runs them in a valid order) and \keyword{minimal} (does not contain unnecessary operations). Information about the purity of operations is used to parallelize the execution plan, improving performance.

% We also propose means to better separate code for value inspection from actual pipeline code: Users can tag inspection code, hide it with filters, and prevent it from running (Sec. \ref{sec:tags-for-code-cells}). Alternatively, they can eliminate it completely from their notebook and instead select context-specific actions on variables to inspect their value (Sec. \ref{sec:introspection}).
We completely eliminate the need for users to \emph{write} value inspection code, by providing context-menu actions on variables to inspect their values (Sec. \ref{sec:introspection}). In an educational context, where value inspection steps must be contained explicitly in the notebook, we propose to tag respective cells, so that they can be eliminated when running a pipeline (Sec. \ref{sec:tags-for-code-cells}). 

% We plan to implement our approach for Safe-DS, a DSL for DS, and evaluate it in a usability study (Sec. \ref{sec:future-plans}). Overall, we expect a positive impact on developer happiness by eliminating bugs introduced by out-of-order execution or stale notebook state and providing a faster feedback loop by running only the code necessary to complete an action.
Overall, we expect that our approach will significantly speed up DS pipeline development, by (1) avoiding bugs resulting from access to stale notebook state, (2) rerunning only the code that must be rerun to update stale state, and (3) eliminating the need to write and understand inspection code. We will verify this claim in a usability study once the prototype for Safe-DS is complete. 

% \todo{
%     \begin{itemize}
%         \item This approach limits the language-agnostic nature of notebooks but we believe the trade-off to be worth it for improved correctness and performance
%     \end{itemize}
% }

\section*{Acknowledgments}

This work was partially funded by the Federal Ministry of Education and Research (BMBF), Germany under the Simple-ML project (grant 01IS18054). We thank the reviewers for their numerous constructive comments and suggestions.

% Bibliography ---------------------------------------------------------------------------------------------------------

% \newpage

\bibliographystyle{IEEEtran}
\bibliography{IEEEabrv,references}

\end{document}